\documentclass[12pt]{iopart}
\usepackage{graphicx}
\usepackage{bm}
\usepackage{iopams}

\newcommand{\g}{pdf}
\usepackage{color}
\usepackage{ulem} 

\newcommand{\removed}[1]{}
\newcommand{\text}[1]{\textrm{#1}}
\newcommand{\eqref}[1]{(\ref{#1})}

\begin{document}

\title[Universal behavior of magnetoresistance in quantum dot arrays]
  {Universal behavior of magnetoresistance in quantum dot arrays with different degree of disorder}

\author{N~P~Stepina$^1$, E~S~Koptev$^1$, A~G~Pogosov$^1$, A~V~Dvurechenskii$^1$,  A~I~Nikiforov$^1$, E~Yu~Zhdanov$^1$ and Y~M~Galperin$^2$ }%
\ead{\mailto{stepina@isp.nsc.ru}}

\address{
$^1$Institute of Semiconductor Physics, 630090 Novosibirsk, Russia
}
\address{$^2$Department of Physics, University of Oslo, P. O. Box
1048 Blindern, 0316 Oslo, Norway  and A. F.  Ioffe Physico-Technical
Institute of Russian Academy of Sciences, 194021 St. Petersburg,
Russia}
\date{\today}

\begin{abstract}
 Magnetoresistance in  two-dimensional array of Ge/Si quantum dots was studied in a wide range of zero-magnetic field conductances, where the transport regime changes from hopping to diffusive one.  The behavior of magnetoresistance is found to be similar for all samples -  it is negative in weak fields and becomes positive with increase of magnetic field.
The result apparently contradicst to existing theories.
 To explain experimental data we suggest that clusters of overlapping quantum dots  are formed.
These clusters are assumed to have metal-like conductance, the
 charge transfer taking place via hopping between the clusters.
Relatively strong magnetic field shrinks electron wave functions decreasing inter-cluster hopping and, therefore, leading to a positive magnetoresistance.
Weak magnetic field acts on ``metallic" clusters destroying interference of electron wave function corresponding to different paths (weak localization) inside clusters.
The interference may be restricted either by inelastic processes, or by the cluster size.
Taking into account WL inside clusters and hopping between them
within  the effective medium approximation we extract effective parameters characterizing charge (magneto) transport.

\end{abstract}

\submitto{\JPC}

\maketitle
\section{Introduction}
This work is aimed at investigation of the (magneto) transport in dense two-dimensional (2D) arrays of self-assembled Ge-in-Si quantum dots  (QDs) grown by molecular-beam epitaxy. Depending on the QD parameters, such arrays demonstrate variety of transport regimes in zero-B field. It has been previously
shown~\cite{Ste10}  that the variation of the (hole) filling factor of QDs,  their structural parameters, density, size and composition leads to essential variation in conductance
(from $10^{-12}$ to $10^{-5}$~Ohm$^{-1}$) and stimulates a crossover from variable range hopping (VRH)~\cite{Shl91} to diffusive transport.

Present work shows that transport properties of a given sample can be governed by interplay between diffusive and hopping contributions.
 Both mechanisms can \textit{simultaneously} determine the transport behavior  of  the QD array in a given sample.
  Interplay of these  types of transport
reveals itself in unusual dependences of the conductance on temperature, $T$, and magnetic field, $B$.  In particular, temperature dependence of zero-$B$ conductance and its magnetic field dependence at large $B$ can be compatible with VRH, while weak-field magnetoresistance (MR) can be rather described by diffusive conductance along the weak localization (WL) concept. To the best of our knowledge, such a combination has been never observed --
usually the observed dependences can be interpreted assuming a single dominant conduction mechanism.

In particular, for hopping conductance theory predicts exponential increase of the conductance with temperature and exponentially large positive MR in high
magnetic fields due to the shrinkage of the localized wave functions in the magnetic field.  Hence, the overlap between localized states
decreases reducing the hopping probability~\cite{Shk84}. In relatively weak magnetic fields and in the VRH regime the
theory predicts a \textit{negative} contribution to MR
associated with suppression by the magnetic field of destructive interference of the wave functions corresponding to electron tunneling between the localized states along different paths~\cite{Ngu85}.   According to the theory, the interference within cigar-shaped domains of impurity states
with the size of the hopping length, $r_h$, may  considerably change the hopping probability leading to a linear in magnetic field $B$ decrease of $\ln [R(B)/R_0]$. Here $R(B)$ is resistance of the sample in magnetic field $B$.

For diffusive transport in the WL regime,
the main mechanism of MR is the interference between different closed-loop trajectories of delocalized electrons experiencing elastic scattering leads to enhanced backscattering~\cite{Ber84}.
The decrease of resistance is due to magnetic-field-induced
phase shift between the wave functions characterizing clockwise and counter-clockwise trajectories, which
suppresses the coherent backscattering and leads to negative MR.
The standard theory of WL is applicable at large conductance,  $G>>G_0=e^2/2\pi^2\hbar$, where $e$ is the electron charge while $\hbar$ is the Plank constant.  In the applicability domain of the WL theory, the quantum contribution should be less than the Drude conductance,
\begin{equation}
G_{\text{D}}=e^2n\tau/m=\pi k_F l G_0.
\end{equation}
Here $n$ and $m$ are the electron density and mass, respectively, $\tau$ is the elastic transport mean free time, $k_F$ and $l$ are the Fermi quasimomentum and the classical mean free path, respectively. Therefore, the  WL theory works well for high values of $k_Fl$. An attempt to extent the WL theory up to  $k_Fl \sim 1$ by including higher corrections in $(k_Fl)^{-1}$ was made by the  authors of Ref.~\cite{Min04}. They reported good agreement between the experiment  and the theory down to conductance values as low as  $10^{-2}G_0$.

Thus, if a particular mechanism dominates the conductance, then  the behavior of MR should be fully determined. The dominant mechanism can be, in turn,
 found from analysis of temperature dependence of zero-$B$ conductance.

We will show that the MR behavior is rather similar  for all our samples
 independently of the magnitudes  and temperature dependences  of  their zero-$B$ resistance.
Namely, for all samples MR is negative in a weak magnetic field and becomes positive as magnetic field increases.
 Moreover, even for the samples with large resistance, the observed negative MR contradicts to the theory~\cite{Ngu85}
based on  account of interference in hopping transport. At the same time, in high magnetic field  MR is positive  even in highly-conductive samples
and its $B$-dependence is similar to that for the hopping mechanism.

 To interpret the observed unusual behavior of MR we suggest that the arrays  contain clusters of close QDs behaving as pieces of diffusive conductor
 embedded in a hopping medium. The resistance of the hopping medium depends on areal density,
  structural parameters and (hole) filling factors of QDs, which determine the localization length, $\xi$, of the hopping problem.
  Diffusive conductance inside the clusters is described by the WL approach assuming that the interference is limited by the shortest
of the two lengths - the phase breaking length, $L_{\phi}$,
and the typical cluster size, $\xi_c$.
  The negative MR in weak magnetic fields is explained by the magnetic-field-induced suppression  of the quantum interference inside the clusters, while the inter-cluster hopping  is only weakly sensitive to the magnetic field.  However, in high magnetic fields it provides a dominant positive contribution to the MR.
 We will develop a semi-quantitative model based on the effective medium approximation (EMA), which allows to evaluate
 both the phase breaking length and the characteristic size of the clusters
for different samples and follow the dynamics between these two lengthes when changing the structural parameters of QDs and temperature.
This picture qualitatively agrees with the observed dependences of resistance both on magnetic field and temperature.

The paper is organized as follows. The samples and the experimental
conditions are described in Sec. II. The analysis of the conductance
versus temperature in samples under study is given in Sec. III. The study of MR is reported in Sec. IV: positive MR (subsection A), negative MR (subsection B).
Our model and interpretation of the observed behaviors of resistance and MR are discussed in Sec.~\ref{Model}.

\section{Samples and experiment}

The samples were grown on a (001) p-Si substrate with  resistivity of 20 $\Omega\cdot$cm by molecular-beam epitaxy of Ge in the
Stranskii-Krastanov growth mode. There were two regimes of growth
allowing to obtain the QDs array with different areal density. In the
first case, the growth temperature for 10 monolayers (ML) Ge layer
was 300$^\circ$C and the growth rate was 0.2 ML/s. As a result, the
areal density of the dots was shown to be $\sim4\times10^{11}$
cm$^{-2}$. In the second case, decrease of the Ge growth temperature down
to 275$^\circ$C with simultaneous increase of the growth rate allows
to reach the twice higher QDs areal density ($\sim8\times10^{11}$cm$^{-2}$).
 To supply holes to the dots, a boron
$\delta$-doped Si layer was inserted 5 nm below the Ge QDs layer.
Because the ionization energy of boron impurities in Si is 45 meV
and the energies of the first ten hole levels in Ge QDs grown at 300$^\circ$C
are 200-400 meV~\cite{a13},  at low temperatures holes leave
impurities  and fill levels in QDs. To get significant changes of
the localization radius $\xi$, the number of holes per dot was
chosen to be 2 and 2.4 for the samples with smaller density. These samples were exposed to additional
annealing in Ar atmosphere during 30 minutes at 550, 575, 600 and
625 $^{\circ}$C. It was proposed that annealing will lead to the smearing of the QD
shape and Ge-Si intermixing increasing overlap  between the hole wave functions belonging to adjacent QDs. The silicon cap layer had a thickness of 40 nm. Metal (Al) source and drain electrodes were deposited on the top of the structure and then heated to 480$^{\circ}$C to form reproducible Ohmic contacts. The resistance along the QD layer was measured by four-terminal method. The temperature stability was controlled using Ge thermometer. The magneto-transport measurements were carried out at 1.3$\div$4.2 K  in magnetic fields 0$\div$10 T.

\section{Temperature dependence of conductivity}

Figure~1 demonstrates the temperature dependences of sheet conductance as Arrhenius plots at zero magnetic field for the low-density samples 2-6 with filling factor $\nu\approx$ 2.4 annealed at different temperatures, for the sample 7 with $\nu\approx$ 2 annealed at 550$^{\circ}$C and for the sample 1 with double density of dots ($\nu\approx$ 3). One can see that the conductance strongly depends on the sample parameters and differs by several orders of magnitude. Within the domain 480$\div$625$^{\circ}$C  of annealing temperatures the conductance increases
with annealing temperature.
\begin{figure}[h]
\centerline{
\includegraphics[width=3.0in]{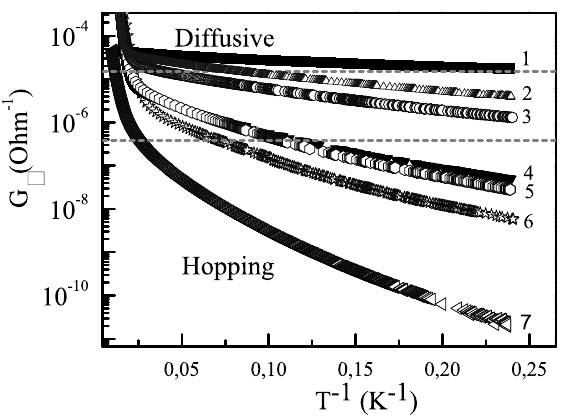}
}
\caption{\label{Fig.1}Temperature dependence of the sheet conductance, $G(T^{-1})$, for the samples with different structural parameters. 1 -- high-density sample with $\nu\sim$~3, 2 -- 6: low-density samples with $\nu\sim$2.4 and different annealing temperatures, $^{\circ}$C: 2 -- 625, 3 -- 600, 4 -- 575, 5 -- 550, 6 -- 480; 7 --  low-density sample with $\nu\sim$2, 550$^{\circ}$C.  }
\end{figure}
It is seen that the conductance of highly conductive samples (1-3) weakly depends on temperature that is typical for diffusive conduction. At the same time, temperature dependence of the conductance  of more resistive samples (4-7) is strong indicating VRH mechanism.
Consequently, for data analysis we used both approaches.

The  temperature dependence of the VRH conductance was specified as
\begin{equation} \label{VRH}
G(T)=\gamma T^m \exp[-(T_0/T)^x]
\end{equation}
where
$\gamma$, $m$ and $T_0$ are material-dependent constants,  $x=1/3$ corresponds to the 2D Mott law while $x=1/2$ corresponds to the Efros-Shklovskii (ES) law.
We
 analyzed the temperature dependence of the
reduced activation energy,
$$w(T)=\partial \ln G(T)/\partial \ln T=m+x(T_0/T)^x,$$
 using the method proposed in Ref.~\cite{Zab84}.
We have found that for
the samples 4--7,
 the temperature dependences of the conductance are well described by the ES law with $x\approx$ 0.5.
From fitted value of $T_0$ we have extracted the localization length as
\begin{equation}\label{xi}
\xi=Ce^2/\varkappa k_BT_0
\end{equation} where
theoretical value of constant $C$ for 2D single-electron hopping is 6.2~\cite{Tsi02}, $\varkappa$ is the
static dielectric constant, $k_B$ is the Boltzmann's constant.
The values of $\xi$ for each sample, calculated using the above mentioned formula, are collected in Table~1.
One can see that the smaller the conductance is the more localized is the system,  as it should be in accordance with the conventional
 scaling theory~\cite{Abr79}.

 More conductive samples (1-3) with  conductance $\sim$$e^2/h$ and weakly dependent  on temperature
cannot be accounted for by  the VRH mechanism.
Therefore, we assume that the mechanism of conductance in those samples has a pronounced diffusive contribution.

In the diffusive regime, the classical contribution to the conductance is temperature-independent and
described by the Drude formula (1).
Quantum contributions can be reliably evaluated when they are small compared with the Drude conductance.
They can be specified as WL corrections, which are due to  interference of elastically-scattered electrons
and corrections induced by electron-electron interaction. The WL corrections
 lead to a logarithmic decrease
of the conductance with decreasing temperature. For 2D case this
correction gives a  negative contribution $\Delta
G_{WL}\approx-G_0\ln (L_\phi/l)$~\cite{Lee85, gor}, where $L_\phi$ is the phase-breaking length, $L_\phi\sim T^\alpha$,
 $\alpha<0$. As a result, $\Delta G_{WL}\propto\ln T$.  The quantum correction due to electron-electron interaction is also proportional to $\ln T$~\cite{alt85}.
Estimates show that for our samples the interaction-induced contribution is less important  than the WL one.

\begin{figure}[h]
\centerline{
\includegraphics[width=3.0in]{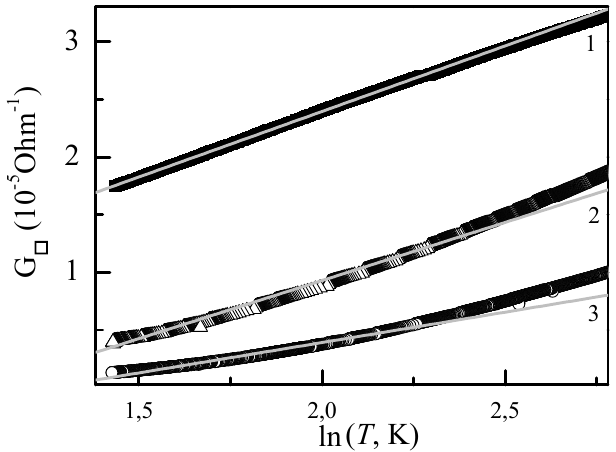}
}
\caption{\label{Fig.2}
Approximation of the conductance of 1-3 samples by the $G\propto \ln T$ law.  Symbols -- experimental results; straight gray lines -- fitting.}
\end{figure}

Figure~\ref{Fig.2} demonstrates fitting the conductances of the 1-3 samples from Fig.~1 by the $G \propto \ln T $ law (gray lines).
 One can see that fitting by this law is really good  only for sample 1; for  samples 2 and 3 some  deviations can be observed.
Previously, we have shown~\cite{Ste10} that at $ G \leq10^{-2}e^2/h$ temperature dependence of the conductance  is well described by the ES law,   whereas at $ G\geq 0.4e^2/h$ the mechanism of the conductance is
close to metallic.
Corresponding boundaries  are shown in Fig.~1 as dashed lines. Based on this classification, we conclude that only the sample 1 shows diffusive behavior of the conductance; the samples 2 and 3 belong to intermediate region, while the samples  4--7 show VRH.

\section{Magnetoresistance of QD array }
\subsection{Positive magnetoresistance}

\begin{figure}[h]
\centerline{
\includegraphics[width=3.0in]{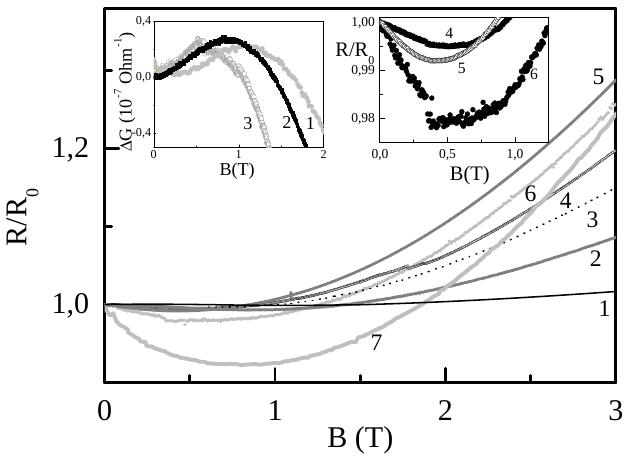}
}
\caption{\label{Fig.3} Magnetoresistance of the same samples as  in Fig.~\ref{Fig.1}.  Insets: enlarged curves   for high-conductance samples (left) and for low-conductance samples (right).}
\end{figure}

Typical dependences of the relative resistance, $R(B)/R_0$, where $R_0(T) \equiv R (B=0,T)$, on transverse magnetic field $B$ at $T=4.2$ K are shown in Fig.~\ref{Fig.3}.
 One can see that all samples demonstrate non-monotonous magnetic field dependences of the resistance, $R$. In low fields $R$ decreases with magnetic field, and then crosses over to an increase. This increase is especially pronounced for high-resistive samples.  We attribute this increase leading to positive MR  to
a shrinkage of the wave functions of localized holes by a transverse magnetic field.

It worth mentioning that pronounced positive MR in high magnetic field is observed even in the samples having diffusive-like conductance in zero magnetic field.
This is very unusual for materials with degenerate charge carriers where classical positive MR vanishes.  Later we will discuss this issue in more detail.

According to Ref.~\cite{Ngu84}, in the case of VRH and
relatively weak magnetic fields this shrinkage results in a quadratic dependence of $\ln R$ on $B$:
\begin{equation} \label{MR1}
\ln[R(B)/R(0)]=(B/B_0)^2.
\end{equation}
Here $B_0=c\hbar\xi^{3/2}_{0}/(s^{1/2}e\xi^{2})$, $\xi_0=(T_0/T)^{1/2}$,
while  $s$ is a numerical coefficient $\approx 10^{-3}$.
According to this expression, $B_0 \propto T^{3/4}$.

Shown in Fig.~4 are the dependences $\ln [R(B)/R_0]$ versus $B^2$ for the sample 5, where conductance
in zero magnetic field is well described by VRH.
\begin{figure}[h]
\centerline{
\includegraphics[width=3.0in]{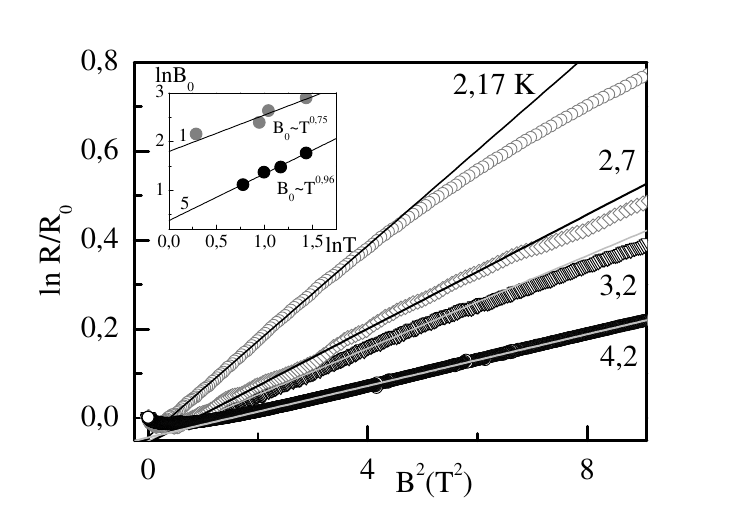}
}
\caption{\label{Fig4} Magnetoresistance  of the sample 5 measured at different temperatures (shown in Kelvins close to the curves). Lines - approximation of the experimental data by Eq.~\eqref{MR1}.  Inset - Temperature dependence of the coefficient $B_0$ for 1 and 5 samples. }
\end{figure}
Despite the fact that
the experimentally measured positive MR is not too large
and log-log plots are not fully reliable, one can notice that
the data collapse if one assumes that $B_0$ is approximately proportional to $T$.
In
 the most conductive sample 1, because of small value of positive MR,
$\lesssim 10\%$, we can approximate
$\ln [R(B)/R_0]$ as  $\Delta R(B)/R_0 \equiv  [R(B)-R_0]/R_0$.
The inset  to Fig.~4 demonstrates the temperature dependence of $B_0$ for the sample 1 and its  approximation by the  $T^{3/4}$ law.
This dependence is typical for the VRH regime though at $B=0$ the conductance of sample 1 is typically diffusive.
We think that in the absence of magnetic field the sample 1 is close to the a metal-to-insulator transition and external magnetic field drives it to the insulating state.
This is, in our view, the reason why the sample having typically diffusive conductance in the absence of magnetic field demonstrates VRH-behavior
in high fields. The classical MR cannot explain the observed positive MR because  in a degenerate gas classical MR is absent  and the ratio between the Fermi energy and temperature in the sample under consideration is rather large.

 Unfortunately, at present time there is no quantitative theory  of the magnetic-field-induced metal-to-insulator  transition.  Nevertheless, we can see that the Eq.~\eqref{MR1} reasonably represents the observed dependences of $\ln [R(B)/R(0)]$ both on magnetic field and temperature.
 Obviously,  for more resistive samples,   the positive MR is larger showing that their behaviors in magnetic field is closer to those typical for  VRH.

\subsection{Negative magnetoresistance}

\begin{figure}[h]
\centerline{
\includegraphics[width=3.0in]{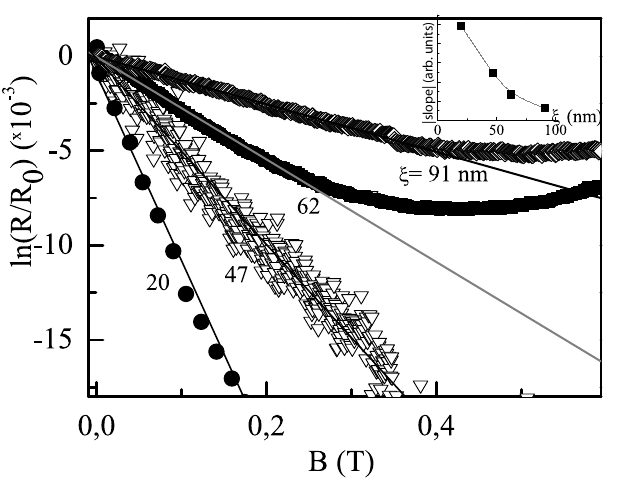}
}
\caption{\label{Fig.5} Logarithm $R(H)/R_0$ versus magnetic field $B$
for the samples 4, 5, 6 and 7 from Fig.1. Numbers at the curves show the values of the localization length obtained from Eq.~\eqref{xi}. Inset shows absolute value of the slope, $|\ln(R/R_0|/R$, as a function of $\xi$.}
\end{figure}

Negative MR in weak magnetic fields was observed in all investigated samples. Insets in
 Fig.~3 show enlarged low-field plots. The left inset shows $\Delta G(B) \equiv G(B)-G(0)$ for high-conductive samples while the right one shows $R(B)/R(0)$ for low-conductive ones. Assuming that conductance of  the low-conductive samples is described by the VRH model, one
is tempted to use the relevant theory of
negative
hopping magnetoresitance, see Ref.~\cite{SpShkl} for a review. According to this theory, logarithm of the ratio $R(H)/R_0$ in relatively week magnetic fields decreases proportionally to $B$~\cite{Ngu85}.  The experimental magnetic field dependences of $\ln(R/R_0)$ versus $B$ for the samples
4-7 are shown in Fig.~\ref{Fig.5}.
Numbers at the curves show the values of the localization length obtained from Eq.~\eqref{xi}.  Though the curves can be fit at low fields by linear functions, it cannot be concluded that the observed magnetoresistance is compatible with the VRH regime.  Firstly, the theory~\cite{SpShkl,Ngu85}
 predicts increase
 of the ratio  $|\ln(R/R_0)|/B$ with localization length.  At the same time, experimentally observed ratio decreases  with $\xi$, see inset in Fig.~\ref{Fig.5}. Secondly, theory~\cite{Sivan88,Entin89} of VRH MR predicts a crossover to quadratic negative MR in very small magnetic field. At the same time, Fig.~\ref{Fig.5} does not show quadratic MR. Finally, the observed negative MR is much smaller that it could be expected for the hopping regime.
The above facts rule out interference of forward paths of hopping carriers~\cite{SpShkl,Ngu85} as the mechanism responsible for observed negative MR.

\section{Model and interpretation} \label{Model}

To explain the negative weak-field MR  observed in all our samples,
as well as the crossover from a negative to a positive MR we propose the model,
based on well known fact of inhomogeneous distribution of self-assembled QDs in the growth plane.

 We assume that our samples with high density of Ge nanoislands contain clusters of closely  located QGs  with diffusive conductance (``metallic droplets") embedded in a host VRH material.
 Then the transport through QD array under study can be represented as a  combination of ``intra-cluster'' diffusive conductance and ``inter-cluster'' hopping one.

If the metallic clusters occupy small areal fraction $\delta$ of the sample,  $\delta < 1/2$, i.e., the system is far enough from the percolation threshold,  the effective conductivity, $\sigma_e$, of a mixture can be found using the effective medium approximation, see Ref.~\cite{Kirk73} for a review.
For a 2D array of circular inclusions with conductivity $G_c$ in the matrix with conductivity $G_h$ the effective medium equation for $G_e$ reads as
\begin{equation} \label{ema1}
\delta \left(\frac{G_c - G_e}{G_c+G_e}\right) + (1-\delta)\left( \frac{G_h - G_e}{G_h+G_e} \right)=0\, .
\end{equation}
The solution of Eq.~(\ref{ema1})  is
\begin{eqnarray} \label{ema}
G_e&=&\frac{1}{2}(1-2\delta)(G_h-G_c)
\nonumber \\ &&
+\frac{1}{2}\sqrt{(G_c+G_h)^2+4\delta(\delta-1)(G_c-G_h)^2}.
\end{eqnarray}

We specify  the hopping conductance in magnetic field as $G_h(B,T)=G_h(0,T)\exp(-B/B_0)^2$ and the conductance of the clusters as $G_c= G_c(0,T)+\Delta G_{WL}$.
The weak localization contribution, $\Delta G_{WL}$, is determined  as
\begin{eqnarray} \label{HLN}
\frac{\Delta G_{WL}}{G_0}&=&\alpha \mathcal{G}, \quad \mathcal{G}(B,L_\phi^*,l)\equiv
       \left[\psi\left(\frac{1}{2}+\frac{\hbar}{2BeL^{*2}_\phi}\right)\right.\nonumber\\
& -&   \psi\left(\frac{1}{2}+\frac{\hbar}{2Bel^2}\right)+\left.2\ln\left(\frac{L^*_\phi}{l}\right)\right].
\end{eqnarray}
Here $\alpha$ is a constant of the order one,
 $\psi(x)$ is the digamma function,
$L_\phi^*$ is an effective phase breaking length, which will be used as an adjustable parameter.
This expression is similar to that obtained by Wittmann \& Schmid~\cite{Witt87} as a modification of the Hikami-Larkin-Nagaoka theory~\cite{Hik80}.  In the original expression, $\alpha =1$ and $L_\phi^*=L_\phi \equiv \sqrt{2D\tau_\phi}$ where $\tau_\phi$ is the phase breaking time, $D$
is the diffusion coefficient.  It was obtained using the diffusion approximation valid for relatively long electron trajectories subjected to interference.  Recently it was shown~\cite{Min03,Min04}  that variety of experimental results can be described by Eq.~(\ref{HLN}), but with renormalized $\alpha$ and $L_\phi^*$, in a much broader domain of conductances than it is required for validity of the diffusion approximation.

 Therefore, we employ Eqs.~\eqref{ema} and (\ref{HLN}) and consider $G_{h,c}(0)$, $B_0$, $\delta$, as well as  $\alpha$, $L_\phi^*$ and the elastic mean free path $l$ as adjustable parameters.
It was found that in the
highly conducting
samples (2-3) with QD areal density  $4\times10^{11}$ cm$^{-2}$, $l\approx 11$ nm  while in the sample 1 with double density of QDs $l\approx 7$ nm.
These values correspond to  typical distances between QD centers in these samples. Therefore, we suggest that the main scattering centers for the delocalized electrons are the QDs. To decrease the number of adjusted parameters we fixed the mean free path $l$ for all the samples with the same areal density as the typical distance between the QDs. At the beginning of approximation, we put $\alpha =1$ and estimate the quantity $B_0$ from positive MR in intermediate magnetic fields where the transport is dominated by VRH. Then we find $G_c,h$ and $\delta$ using the effective medium approximation \eqref{ema}. The parameters relevant to WL are found
putting $G_c= G_c(0,T)+\Delta G_{WL}$ and then  using Eq.~\eqref{HLN}. After that we modify $\alpha$ and then repeat the procedure to get the best fit. It turns out that the best fit for the more conductive samples (1-4) is achieved for $\alpha =1$.

\begin{figure}
\centerline{
\includegraphics[width=3.0in]{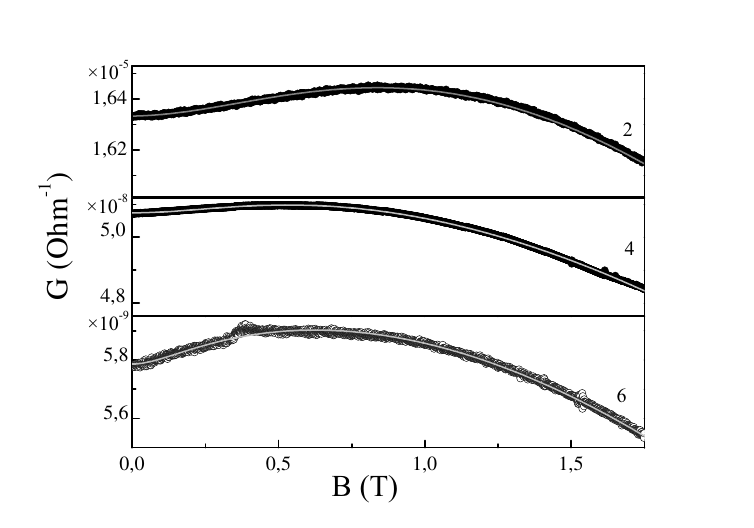}
}
\caption{\label{fig7} Approximation of the experimental data for samples 2, 4 and 6 from Fig.~1 by Eq.~\eqref{ema}}.
\end{figure}

Shown in Fig.~\ref{fig7}
are examples of fitting of the experimental data for the samples 2, 4 and 6 by Eqs.~(\ref{ema}) and \eqref{HLN}. The extracted parameters for all the samples  are collected in Table~1.

\begin{table}[h]
\centerline{
\begin{tabular}{cr l c  c l rl}
\hline\noalign{\smallskip}
$N$ & $\xi$\phantom{0} &\phantom{00}  $\alpha$ & $L_\phi^*$  & $G_h(0)$ & \phantom{0}$G_c(0)$ & $L_\phi $\phantom{0}& \phantom{0}$\delta$\\
\hline \noalign{\smallskip}
1&  &1&41&1.7$\times 10^{-5}$&1.8$\times 10^{-5}$&46&0.03\\
2&  &1&32&1.6$\times 10^{-5}$&2.0$\times 10^{-5}$&37&0.15\\
3& &1&41&5.8$\times 10^{-7}$&2.0$\times 10^{-5}$&56&0.31\\
4&91&1&43&4.4$\times 10^{-9}$&9.4$\times 10^{-6}$&58&0.46\\
5&62&0.017&62&8.7$\times 10^{-9}$&6.2$\times 10^{-7}$&270&0.37\\
6&47&0.0014&62&2.8$\times 10^{-9}$&4.6$\times 10^{-8}$&1000&0.31\\
7&21&0.00014&65&1.6$\times 10^{-10}$&2.1$\times 10^{-9}$&5000&0.30\\
\hline \noalign{\smallskip}
\end{tabular}}
\caption{\label{opt}Characteristic parameters of the samples for 4.2K obtained from Eqs.~\eqref{xi}, \eqref{ema} and \eqref{HLN}. Lengths are given in nm, conductances -- in (Ohm)$^{-1}$.}
\end{table}

We conclude that the proposed model provides a reasonable interpretation of the experimental findings for the samples 3-5 where the partial
conductances $G_h$ and $G_c$ are essentially different making EMA efficient. For these samples
 we are able to
(quantitatively)
reproduce non-monotonous dependence of resistance on magnetic field.
   The extracted adjustable parameters $\delta, G_{h,c}, \alpha$  have reasonable magnitudes compatible with the initial assumptions.

For highly-conductive samples (1,2), our fitting provides low values of the fraction $\delta$ of metallic clusters.  In addition, fitted values of $G_c$ and $G_h$ in zero magnetic field are close to each other. These results apparently contradict to the concept of hopping between relative rare metallic clusters.
However, in high magnetic fields these samples show noticeable positive MR, which cannot be interpreted by the classical theory
of metallic conductance. We think that magnetic-field-induced shrinking of holes' wave functions leads to a decrease of their overlap between different clusters.
As a result, the magnetic field drives the sample into insulating regime -- the regions of diffusive hole transport become  connected by hopping regions, which are responsible for observed positive MR. Unfortunately, the crossover region cannot
quantitatively accounted
by our simple model.
We are not aware of any quantitative theory of such crossover.
Therefore,  behaviors of the resistance of samples 1 and 2 are interpreted only qualitatively.

For highly-resistive samples (5-7) our procedure leads to  low values of the adjustable parameter $\alpha$ for highly resistive samples.
The fitted value of $\alpha$ decreases with decrease of conductance. A similar trend was observed in Ref.~\cite{Min04}, where this behavior has been attributed to a crossover  between WL and so-called weak insulator (WI) regime.  The WI regime is the case close to the metal-to-insulator transition where $k_F l \gtrsim 1$ and $G(0) \lesssim G_0$.  A special name for this regime was introduced because the MR is still well described by the WL expression \eqref{HLN} with reduced pre-factor $\alpha$
and fitted parameter $L_\phi^* \lesssim L_\phi$~\cite{Min07}.
In particular, the authors of Ref.~\cite{Min04} have shown that this approach is compatible with experimental results for disordered 2D GaAs/In$_x$Ga$_{1-x}$As/GaAs quantum well structures down to
 $G \gtrsim 10^{-2}G_0$.

 The qualitative
conclusion that $\alpha$ should decrease with decrease of conductance is
compatible with
the theory~\cite{Aleiner99, Min04} showing that if the two-loop localization correction as well as the interplay of the weak localization and the interaction are taken into account.
However, the  theory does not allow for  very small values of $\alpha$ extracted for the samples 6-7 and  the above method of extracting parameters for these samples with high resistance can be considered
 as an empirical one.

When fitting their result, authors of Ref.~\cite{Min07} assumed the  cut-off length $L_\phi^*$ is given by the empirical expression
\begin{equation} \label{lstar}
1/L^{*2}_\phi =1/L_{\phi}^{2}+ 1/\xi^{*2}\, .
\end{equation}
where   $L_{\phi}$  is  the dephasing length and $\xi^*$ is the  2D localization length~\cite{Lee85},
\begin{equation} \label{lr}
\xi^*\approx l\exp(\pi k_Fl/2)\, .
\end{equation}
\

We also use the interpolation~\eqref{lstar} assuming that there exists a characteristic cut-off length $\xi_c$, which limits the size of interfering trajectories.
This length can be ascribed to characteristic  size of clusters with diffusive conductance formed on our samples.
We think that charge transport occurs via hopping between such clusters.
Having this mechanism in mind we replace $\xi^* \rightarrow \xi_c$ in Eq.~\eqref{lstar}.

At low temperatures the dephasing length increases, and the fitted value $L^*_\phi$ should saturate at the typical cluster size, $\xi_c$.
Interestingly, the values of $\xi_c \sim 60-80$~nm, compatible with our experimental results for all the samples, have the same order of magnitude as the localization length $\xi^*$, which can be roughly estimated from the density of QDs using  Eq.~\eqref{lr}.  The ``hopping" localization length, $\xi$,  given by Eq.~\eqref{xi}  has also the same order of magnitude.
At $L_{\phi} \lesssim \xi_c$ temperature dependence of $L_\phi$ should manifest itself through temperature dependence of the negative MR. We will address this issue in the next section.

\subsection{Temperature dependence of the phase breaking length} \label{Sec4c}
\begin{figure}[h]
\centerline{
\includegraphics[width=3.0in]{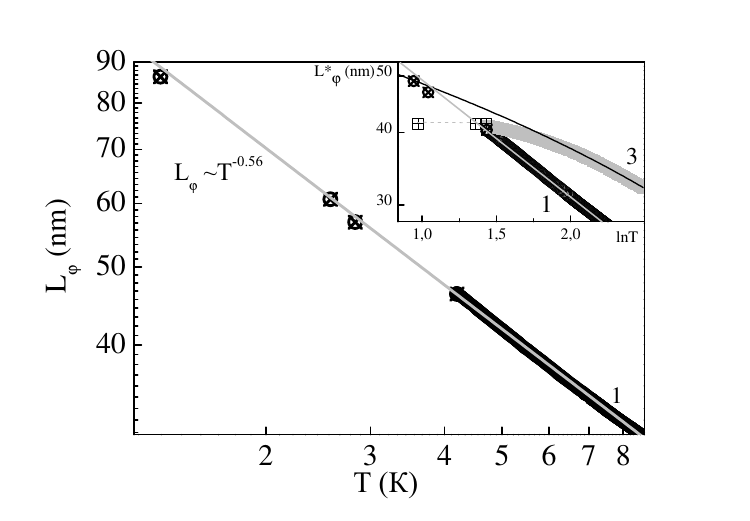}
}
\caption{ $L_{\phi}(T)$ dependence, restored from $G(T)$ data for sample 1 (close symbols).
 Open symbols - the result of analysis of MR measured at low temperatures  for the same sample.
  Line - approximation of $L_{\phi}(T)$ dependence. Inset - $L^*_{\phi}(T)$ plot for samples 1 and 3.  \label{fig72}}
\end{figure}

As the conductance of the samples is determined by a competition between ``intra-cluster'' and ``inter-cluster'' contributions, its temperature dependence
depends on
their
 interplay.
 For most resistive samples (4-7) temperature dependence of zero-field conductance is dominated by the ``inter-cluster" hopping resulting in the Efros-Shklovskii law.
For more conductive samples zero-$B$ $G(T)$ dependence can be approximated by a logarithmic law  that is compatible with the WL mechanism.  Assuming this mechanism one can ascribe the $G(T)$-dependence to the temperature dependence of the dephasing length $L_\phi$. This dependence can be extracted in two ways: (i)  from  temperature dependence of the conductance,  and (ii)  from the MR-data  measured at different temperatures. In both procedures we actually extract the value of $L_\phi^*$ rather that ``true" $L_\phi$, which is determined by inelastic processes.
As it follows from Eq.~\eqref{lstar}
 at  $L_\phi \ll \xi_c$ the effective length  $L^*_\phi$  coincides with the dephasing length $L_\phi$. In the opposite case
$L^*_\phi \approx \xi_c$, and it is temperature-independent since the interference area is limited by the cluster size.
Even if  $L_\phi < \xi_c$ at relatively high temperature, this relationship can be reversed at low temperatures since $L_\phi$ is a decreasing function of temperature. Therefore, temperature dependence of $L^*_\phi$  saturates at low temperatures. Temperature-independent $L^*_\phi$ is also characteristic for the samples with large structure disorder where $L_\phi \gtrsim \xi_c$ at any accessible temperature.

Shown in Fig.~\ref{fig72} is the $L_{\phi}(T)$ dependence for the most conductive sample 1, restored from the $G(T)$ data  (closed symbols) and from the MR data (open symbols) using Eqs.~\eqref{lstar} and \eqref{lr}.   Gray line is the approximation of the experimental data by $L_{\phi}\approx T^{-0.56}$; the exponent 0.5 is typical for the dephasing due to electron-electron scattering in 2D systems~\cite{Ale02}. This behavior agrees  with the results of MR measurements, according to which  for the 1 sample $L_{\phi}<\xi_c$ at 4.2 K. Obviously, the $L_{\phi}$ values obtained from the MR measurements at low temperature (open symbols) roughly extend the approximating line. The $L_\phi^* (T)$ dependences for the samples 1 and 3 extracted from the temperature dependence of conductance (closed symbols) and MR data (open symbols), are shown in the inset. One can see a clear tendency to saturation at low temperatures for the sample 3.
Although for this sample  the inequality $L_{\phi}<\xi_c$ holds, it is not very strong and
 the temperature dependence of $L^*_{\phi}$ extracted from fitting experimental results tends to saturation.
Therefore, $L_{\phi}(T)$-dependence
extracted following our procedure
determines  the behavior of the conductance in all investigated temperature range  only for the sample~1.

\section{Conclusion}

We have
studied conductance of dense 2D arrays of Ge-in-Si quantum dots  versus temperature and magnetic field.  The set of samples was obtained by different  annealing procedures and growth regimes  resulting in very different  values of resistance and quantitatively different  temperature dependences. However, we found that magneto-transport was rather universal: all the samples showed negative MR in weak magnetic fields, which then  crossed over to a positive MR in high magnetic fields.
To interpret the results we have proposed a  model describing
dependences of resistance of dense  QD arrays both on temperature and magnetic field.  Based on fitting of the experimental data by our model we  confirm that
 the samples contain clusters of closely located QDs with diffusive intra-cluster conductance.
These clusters are embedded in a host VRH material
responsible for the inter-cluster transport strongly dependent on the QDs structural parameters and their filling with carriers. Annealing of the samples provokes  a reconstruction of the QDs leading to increase of the wave function overlapping and increase of the conduction inside clusters as well between them.   Resulting conductance of the structure is determined by a competition between intra- and inter-cluster contributions, which we describe in terms of the effective-medium approximation.
Regarding magneto-transport, we
 conclude that positive magneto-resistance  in high magnetic field is a result of suppression of the conductance of the hopping matrix due to shrinkage of the holes' wave functions, whereas negative weak-field MR is due to weak localization contributions to the conductance of the clusters.  We were able to extract relevant parameters for the model, which facilitates description of the crossover from definitely insulating (VRH) behavior to a quasi-metallic one.

\ack{
This work is done with support of RFBR (Grant 13-02-00901) and ERA-NET-SBRAS. Authors are thankful to G. M. Minkov, A. V. Germanenko and A. V. Nenashev for very useful discussions.
}

\end{document}